\documentclass[fleqn,showpacs,aps,graphicx]{revtex4}
\usepackage{graphicx}
\usepackage{mathrsfs}
\usepackage{amsfonts}
\usepackage{amssymb}
\usepackage[mathscr]{eucal}
\usepackage{amssymb}
\usepackage{amsmath}
\usepackage{dcolumn}
\usepackage{bm}

\begin{document}

\title{Efficient single-photon entanglement concentration for quantum communications}

\author{Lan Zhou$^{1,2}$}
\address{$^1$ College of Mathematics \& Physics, Nanjing University of Posts and Telecommunications, Nanjing,
210003, China\\
$^2$Key Lab of Broadband Wireless Communication and Sensor Network
 Technology, Nanjing University of Posts and Telecommunications, Ministry of
 Education, Nanjing, 210003, China\\}

\date{\today }

\begin{abstract}

We present two protocols for the single-photon entanglement concentration.
 With the help of the 50:50 beam splitter, variable beam splitter and an
auxiliary photon, we can concentrate a less-entangled single-photon
state into a maximally single-photon entangled state with some probability.
The first protocol is implemented with linear optics and the second protocol
is implemented with the cross-Kerr nonlinearity. Our two protocols do not need two pairs of entangled states shared by the
two parties, which makes our protocols
more economic. Especially, in the second protocol, with the help of the cross-Kerr nonlinearity,
 the sophisticated single photon detector is not required. Moreover, the second protocol
 can be reused to get higher success probability. All these advantages may make
 our protocols  useful in the long-distance quantum communication.
\end{abstract}
 \maketitle

\section{Introduction}
Entanglement plays an important role in the current quantum information processing \cite{book,rmp}.
Most  quantum information protocols such as quantum teleportation \cite{teleportation1}, quantum dense coding \cite{densecoding1}, quantum state sharing \cite{QSS1,QSS2,QSS3} and
other protocols \cite{Ekert91,BBM92,long,two-step,lixhpra,QKDdeng1}  need the entanglement  to set up the quantum channel.
Among all the entanglement forms, the single-particle entanglement with the form
$\frac{1}{\sqrt{2}}(|0,1\rangle_{AB}+|1,0\rangle_{AB})$  may
be the simplest one. Here $|0\rangle$ and $|1\rangle$
mean none particle and one particle, respectively, and the subscripts $A$ and $B$ mean different locations. The single-photon entangled state corresponds to a superposition state in which the single photon is in two different locations $A$ and $B$.
 In 2002, Lombardi
\emph{et al.} reported their experimental results for teleportating a vacuum-one-photon qubit
with the fidelity of 0.953 \cite{lombardi}. In 2005, Chou \emph{et al.}
    realized the experiment for observing the entanglement between
two atomic ensembles located in distant, spatially separated set-ups.
It is essentially the creation of entanglement by storing the single-photon
entanglement into the atomic-ensemble-based quantum memory \cite{chou}.
The most important application of the single-photon entanglement may be the quantum repeater protocol
in long-distance quantum communication. For example, in the well known Duan-Lukin-Cirac-Zoller
(DLCZ) repeater protocol \cite{memory,singlephotonrepeater3}, with one pair source and one quantum memory
at each location, the quantum repeater can entangle two remote locations A and B. It can be written as
$\frac{1}{\sqrt{2}}(|e\rangle_{A}|g\rangle_{B}+|g\rangle_{A}|e\rangle_{B})$. The $|e\rangle$ and $|g\rangle$ represent
the excited state and the ground state of the atomic ensembles, respectively. Recently, Gottesman \emph{et al.} proposed a protocol for building an interferometric telescope based on the single-photon entangledstate\cite{telescope}. With the help of the single-photon entanglement, the protocol has the potential to eliminate the baseline length limit, and allows in principle the interferometers with arbitrarily
long baselines.

Unfortunately, in the practical transmission, similar to the other types of entanglement,
 the noisy environment and the imperfect operation may make the maximally single-photon entangled state degrade into a mixed state or become a pure less-entangled state. For example,  the relative phase between
 the different spatial modes is sensitive to the path-length instabilities \cite{repeater1}.
  It has become an inherent drawback in quantum repeaters which can make the long-distance quantum communication extremely
  difficult. In the single-photon entanglement, the sensitive phase fluctuation
  cause a phase-flip error on the maximally entangled state, which will make
 the  single-photon entangled state
   $\frac{1}{\sqrt{2}}(|0,1\rangle_{AB}+|1,0\rangle_{AB})$  become $\frac{1}{\sqrt{2}}(|0,1\rangle_{AB}-|1,0\rangle_{AB})$.   In 2008, the group of Gisin proposed an effective protocol for the purification of single-photon entanglement with linear optics \cite{sangouard}, which was realized in experiment by themselves in 2010 \cite{sangouard2}.
    On the other hand, as pointed out by Ref. \cite{memory}, as we cannot ensure the pair sources which are excited by the
  synchronized classical pumping pulses have the same probability to generate the single photon, after the entanglement generation, it may also lead the less-entangled state with the form of $\alpha|0,1\rangle_{AB}+\beta|1,0\rangle_{AB}$, with $|\alpha|^{2}+|\beta|^{2}=1$. In the entanglement connection stage, if we use  such less-entangled states $\alpha|0,1\rangle_{AB}+\beta|1,0\rangle_{AB}$ and $\alpha|0,1\rangle_{CD}+\beta|1,0\rangle_{CD}$ to connect the entanglement with the entanglement swapping, we will obtain the lesser
   quality quantum entanglement channel $\alpha^{2}|0,1\rangle_{AD}+\beta^{2}|1,0\rangle_{AD}$\cite{memory,shengqic}. So before entanglement connection, we need to recover the less-entangled state into the maximally entangled state.

  Entanglement concentration is a powerful way to recover the pure less-entangled state into a maximally entangled
  state probabilistically \cite{shengqic,C.H.Bennett2,swapping1,swapping2,zhao1,Yamamoto1,wangxb,shengpra2,shengpra3,shengpra4,dengpra,wangc}. Most  entanglement concentration protocols (ECPs) are focused on the two-particle entanglement, such as the Schimidit projection method \cite{C.H.Bennett2}, the ECP based on the entanglement swapping \cite{swapping1},
 the ECP using unitary transformation \cite{swapping2},
  and the ECPs with linear optics and  cross-Kerr nonlinearity \cite{shengqic,zhao1,Yamamoto1,wangxb,shengpra2,shengpra3,dengpra}. In 2009, we have proposed an ECP for single-photon entanglement
  with the help of cross-Kerr nonlinearity \cite{shengqic}. In that protocol, after the two parties Alice and Bob pick up the successful case, they should send the photon in the spatial mode $a_{2}b_{2}$ to make a collective measurement. Moreover, during each concentration step, they require two pairs of nonlocal single-photon entanglement states and after the measurement at least one pair of entangled state can be remained. So it is not strange that the protocol has higher efficiency than others. Though the protocol of Ref. \cite{shengqic} can reach a higher success probability than the protocol of Ref. \cite{zhao1}, it is not an optimal one. The main reason is that the protocol is not based on the local operation and classical communication (LOCC). On the other hand, in order to obtain a higher success probability, it requires the
  phase shift of the coherent state to reach a large value of $\pi$ in the single-photon level. However, natural cross-Kerr nonlinearities are extremely weak, which makes it difficult to realize the protocol under current experimental conditions \cite{kok1,kok2}.

  In this paper, we present two efficient ECPs for less-entangled single-photon state with local single photon. Both protocols only require one
  less-entangled single-photon  state and a conventional single photon. In the first protocol, we use the linear optical
  elements to complete the task and it can reach the same success probability as Ref. \cite{zhao1}. In the second protocol, we adopt the
  weak cross-Kerr nonlinearity to improve the first ECP. The second protocol can get a higher success probability as it can be reused to further concentrate the discarded items in the first protocol. This advantage makes it more feasible
  than others.  Certainly, in order to perform both the protocols successfully, we should know the initial coefficients $\alpha$ and $\beta$ to
   prepare the state of single photon. Actually, in the previous ECPs in Refs.\cite{swapping2,shengpra3,shengpra4,dengpra,wangc}, they also need to the initial coefficients. In a practical operation, one can measure the enough samples to obtain the exact values of  $\alpha$ and $\beta$ \cite{dengpra}.

   This paper is organized as follows: In Sec. II, we present our first ECP with linear optics. In Sec. III, we describe
  the second ECP with weak cross-Kerr nonlinearity. In Sec. IV, we briefly discuss the future experiment, calculate the total success probability  and make a  conclusion.

\section{Single-photon entanglement concentration with linear optics}

 \begin{figure}[!h]
\begin{center}
\includegraphics[width=8cm,angle=0]{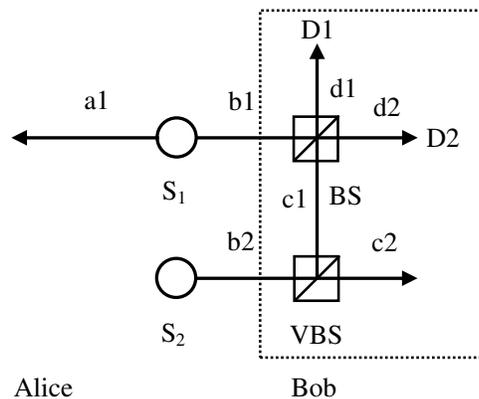}
\caption{A schematic drawing of our single-photon entanglement concentration protocol with linear optics. It is constructed by a
50:50 beam splitter (BS) and a variable beam splitter (VBS). Alice and Bob share a less-entangled single-photon state in the mode $a_{1}$ and $b_{1}$. Another single photon source emits an auxiliary photon in the mode $b_{2}$. The BS is located in the middle of Alice and Bob and it is used to couple the mode $a_{1}$ and $c_{2}$. The VBS is used to adjust the coefficients of the entangle state between the mode $a_{1}$ and $c_{2}$, and finally to attain the maximally entangled state. }
\end{center}
\end{figure}

  The basic principle of our first ECP is shown in Fig. 1. We suppose that the single photon source S$_{1}$ emits a photon and sends it to Alice and Bob in the spatial modes a$_{1}$ and b$_{1}$, which can create a less-entangled single-photon  state $|\phi_{1}\rangle_{a_{1}b_{1}}$. $|\phi_{1}\rangle_{a_{1}b_{1}}$ can be written as
 \begin{eqnarray}
|\phi_{1}\rangle_{a_{1}b_{1}}&=&\alpha|1,0\rangle_{a_{1}b_{1}}+\beta|0,1\rangle_{a_{1}b_{1}},\label{singlephoton}
 \end{eqnarray}
where $\alpha$ and $\beta$ are the coefficients of the initial entangled state, $|\alpha|^{2}+|\beta|^{2}=1$ and $\alpha\neq\beta$.

  Then another single photon source S$_{2}$ emits an auxiliary photon and sends it to Bob in the spatial mode b$_{2}$. Bob makes this auxiliary photon pass through a variable beam splitter (VBS) with the transmission of t, which can create another single-photon entangled state between the spatial mode c$_{1}$ and c$_{2}$ of the form
\begin{eqnarray}
|\phi_{2}\rangle_{c_{1}c_{2}}&=&\sqrt{1-t}|1,0\rangle_{c_{1}c_{2}}+\sqrt{t}|0,1\rangle_{c_{1}c_{2}}.\label{VBS}
 \end{eqnarray}
In this way, the state of the whole two-photon system can be written as
\begin{eqnarray}
&&|\phi\rangle_{a_{1}b_{1}c_{1}c_{2}}=|\phi_{1}\rangle_{a_{1}b_{1}}\otimes|\phi_{2}\rangle_{c_{1}c_{2}}\nonumber\\
&=&\alpha\sqrt{1-t}|1,0,1,0\rangle_{a_{1}b_{1}c_{1}c_{2}}+\beta\sqrt{t}|0,1,0,1\rangle_{a_{1}b_{1}c_{1}c_{2}}\nonumber\\
&+&\alpha\sqrt{t}|1,0,0,1\rangle_{a_{1}b_{1}c_{1}c_{2}}+\beta\sqrt{1-t}|0,1,1,0\rangle_{a_{1}b_{1}c_{1}c_{2}}.\label{entangle generation}
\end{eqnarray}
Then, Alice and Bob make the photons in the spatial modes b$_{1}$ and c$_{1}$ enter the $50:50$ beam splitter (BS), which can make
\begin{eqnarray}
\hat{b}_{1}^{\dagger}|0\rangle=\frac{1}{\sqrt{2}}(\hat{d}_{1}^{\dagger}|0\rangle-\hat{d}_{2}^{\dagger}|0\rangle)\nonumber\\ \hat{c}_{1}^{\dagger}|0\rangle=\frac{1}{\sqrt{2}}(\hat{d}_{1}^{\dagger}|0\rangle+\hat{d}_{2}^{\dagger}|0\rangle).
\end{eqnarray}

Here, the $\hat{b}_{j}^{\dagger}$, $\hat{c}_{j}^{\dagger}$ and $\hat{d}_{j}^{\dagger}$ with $j=1,2$ are the creation operators for the spatial mode $b_{j}$, $c_{j}$ and $d_{j}$, respectively.   After passing through the BS, the whole two-photon system can evolve to
\begin{eqnarray}
|\phi\rangle_{a_{1}d_{1}d_{2}c_{2}}&=&\frac{\alpha\sqrt{1-t}}{\sqrt{2}}|1,1,0,0\rangle_{a_{1}d_{1}d_{2}c_{2}}+\frac{\beta\sqrt{t}}{\sqrt{2}}|0,1,0,1\rangle_{a_{1}d_{1}d_{2}c_{2}}\nonumber\\
&+&\frac{\alpha\sqrt{1-t}}{\sqrt{2}}|1,0,1,0\rangle_{a_{1}d_{1}d_{2}c_{2}}-\frac{\beta\sqrt{t}}{\sqrt{2}}|0,0,1,1\rangle_{a_{1}d_{1}d_{2}c_{2}}\nonumber\\
&+&\alpha\sqrt{t}|1,0,0,1\rangle_{a_{1}d_{1}d_{2}c_{2}}+\frac{\beta\sqrt{1-t}}{\sqrt{2}}|0,2,0,0\rangle_{a_{1}d_{1}d_{2}c_{2}}\nonumber\\
&-&\frac{\beta\sqrt{1-t}}{\sqrt{2}}|0,0,2,0\rangle_{a_{1}d_{1}d_{2}c_{2}}.\label{BS}
\end{eqnarray}

According to Eq. (\ref{BS}), it can be easily found that if they pick up the case that the detector D$_{1}$ clicks exactly one photon, Eq. (\ref{BS}) will collapse to
\begin{eqnarray}
|\phi\rangle_{a_{1}c_{2}}&=&\alpha\sqrt{1-t}|1,0\rangle_{a_{1}c_{2}}+\beta\sqrt{t}|0,1\rangle_{a_{1}c_{2}},\label{connection}
\end{eqnarray}
while if they pick up the case that the detector D$_{2}$ clicks exactly one photon, Eq. ({\ref{BS}}) will collapse to
\begin{eqnarray}
|\phi'\rangle_{a_{1}c_{2}}&=&\alpha\sqrt{1-t}|1,0\rangle_{a_{1}c_{2}}-\beta\sqrt{t}|0,1\rangle_{a_{1}c_{2}}.\label{connection1}
\end{eqnarray}
There is only a phase difference between the Eq. (\ref{connection1}) and Eq. (\ref{connection}). Alice or Bob only need to perform a phase flip operation with the help of a
half-wave plate, then Eq. (\ref{connection1}) can be easily converted to Eq. (\ref{connection}).
 According to  Eq. (\ref{connection}), we can easily find that if a suitable VBS can be provided, which makes $t=\alpha^{2}$, Alice and Bob can make the coefficients $\alpha\sqrt{1-t}=\beta\sqrt{t}$. In this case,  Eq. (\ref{connection}) can evolve to
\begin{eqnarray}
|\Phi\rangle_{a_{1}c_{2}}&=&\frac{1}{\sqrt{2}}(|1,0\rangle_{a_{1}c_{2}}+|0,1\rangle_{a_{1}c_{2}}).\label{max entangle}
\end{eqnarray}
It is obvious that Eq. ({\ref{max entangle}}) is the maximally entangled state between the modes a$_{1}$ and c$_{2}$. That is to say, in the protocol, with the help of the Bell measurement and the suitable VBS with the transmission $t=\alpha^{2}$,
Alice and Bob can distill a maximally single-photon entangled state from the arbitrary less-entangled single-photon state, with the success probability of 2$|\alpha\beta|^{2}$.

 \section{Single-photon entanglement concentration with weak cross-Kerr nonlinearity}

So far, we have explained the first ECP with linear optics. According to the measurement results of the detectors, they can ultimately obtain
the maximally entangled state. However, the whole protocol is not an optimal one.
 \begin{figure}[!h]
\begin{center}
\includegraphics[width=8cm,angle=0]{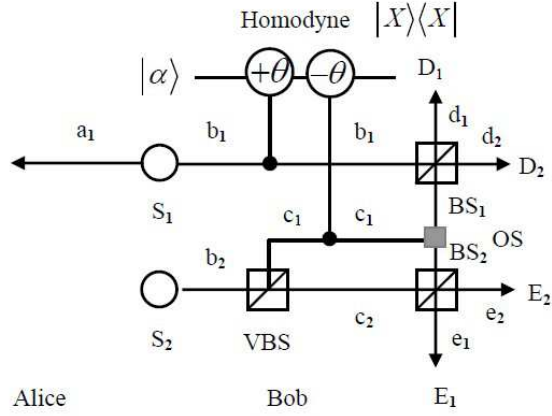}
\caption{A schematic drawing of our single-photon ECP with weak cross-Kerr nonlinearity. Two cross-Kerr nonlinearities are adopted to distinguish the $a_{1}^{\dagger}c_{1}^{\dagger}|0\rangle$ and $b_{1}^{\dagger}c_{2}^{\dagger}|0\rangle$ from $a_{1}^{\dagger}c_{2}^{\dagger}|0\rangle$ and $b_{1}^{\dagger}c_{1}^{\dagger}|0\rangle$. The optical switch (OS) can lead the photon in $c_{1}$ mode into the BS$_{1}$ or BS$_{2}$
 according to the different phase shift. The VBS is used to adjust the coefficients of the entangle state between the mode $a_{1}$ and $c_{2}$, and finally to attain the maximally entangled state. }
\end{center}
\end{figure}
In our second ECP, we adopt the cross-Kerr nonlinearity to complete the task. We will prove that this ECP can be repeated to obtain a higher success probability. Before explaining our ECP, we would like to briefly introduce the cross-Kerr nonlinearity. The cross-Kerr nonlinearity has been widely studied in the quantum information field, such as the quantum state preparation and detection \cite{detection1}, the implementation of quantum logic gates \cite{gate,gate1},  Bell-state analysis \cite{QND2,shengbellstateanalysis}, and so on \cite{qi,he1,he2,he3,lin1,lin2,lin3,xiayan1,xiayan2,zhangzhiming,guoyu}.
The cross-Kerr nonlinearity can be usually described with its Hamiltonian \cite{gate}
\begin{eqnarray}
H_{ck}=\hbar\chi\hat{n_{a}}\hat{n_{b}},
\end{eqnarray}
where $\hbar\chi$ is the coupling strength of the nonlinearity, which depends on the cross-Kerr-material. $\hat{n_{a}}$ and $\hat{n_{b}}$ are the photon number operators for the mode a and mode b. In the practical application, a coherent beam in the state $\alpha_{p}$ and a single photon with the form $|\psi\rangle=\gamma|0\rangle+\delta|1\rangle$ interact with the cross-Kerr material. After the interaction, the system can evolve to
\begin{eqnarray}
U_{ck}|\psi|\alpha\rangle&=&(\gamma|0\rangle+\delta|1\rangle)|\alpha\rangle
\rightarrow\gamma|0\rangle|\alpha\rangle+\delta|1\rangle|\alpha e^{i\theta}\rangle.\label{QND}
\end{eqnarray}
Here, $|0\rangle$ and $|1\rangle$ mean none photon and one photon, respectively. $\theta=\chi t$, and t means the interaction
time for the signal with the nonlinear material. According to Eq. (\ref{QND}), we can find that the phase shift of the coherent state is
directly proportional to the photon number.

As shown in Fig. 2, in our second single-photon ECP, we introduce the weak cross-Kerr nonlinearity to construct the  quantum nondemolition  detector (QND). As described in Section. II, with the help of the VBS, the whole two-photon system can be described by the Eq. (\ref{entangle generation}). Then, Bob makes the photons in the spatial modes b$_{1}$ and c$_{1}$ pass through the QND. After the homodyne measurement, the whole two-photon system combined with the coherent state can be written as
\begin{eqnarray}
|\phi_{1}\rangle_{a_{1}b_{1}}&\otimes&|\phi_{2}\rangle_{c_{1}c_{2}}\otimes|\alpha\rangle
\rightarrow\alpha\sqrt{1-t}|1,0,1,0\rangle_{a_{1}b_{1}c_{1}c_{2}}|\alpha e^{-i\theta}\rangle\nonumber\\
&+&\beta\sqrt{t}|0,1,0,1\rangle_{a_{1}b_{1}c_{1}c_{2}}|\alpha e^{i\theta}\rangle
+\alpha\sqrt{t}|1,0,0,1\rangle_{a_{1}b_{1}c_{1}c_{2}}|\alpha\rangle\nonumber\\
&+&\beta\sqrt{1-t}|0,1,1,0\rangle_{a_{1}b_{1}c_{1}c_{2}}|\alpha\rangle.\label{QND1}
\end{eqnarray}

It can be seen that the items $\alpha\sqrt{1-t}|1,0,1,0\rangle_{a_{1}b_{1}c_{1}c_{2}}$ and $\beta\sqrt{t}|0,1,0,1\rangle_{a_{1}b_{1}c_{1}c_{2}}$ can lead the coherent state pick up the phase shift of $-\theta$ and $\theta$, respectively, while both the items $\alpha\sqrt{t}|1,0,0,1\rangle_{a_{1}b_{1}c_{1}c_{2}}$ and $\beta\sqrt{1-t}|0,1,1,0\rangle_{a_{1}b_{1}c_{1}c_{2}}$ lead coherent state pick up no  phase shift. As in the practical measurement, the phase shift $-\theta$ and $\theta$ are undistinguishable \cite{gate}, Alice and Bob only need to select the items corresponding to the phase shift $\theta$(-$\theta$) and discard the other items. Then the Eq. (\ref{QND1}) will evolve to
\begin{eqnarray}
|\psi_{1}\rangle_{a_{1}b_{1}c_{1}c_{2}}&=&\alpha\sqrt{1-t}|1,0,1,0\rangle_{a_{1}b_{1}c_{1}c_{2}}
+\beta\sqrt{t}|0,1,0,1\rangle_{a_{1}b_{1}c_{1}c_{2}},
\label{after QND}
\end{eqnarray}
with the success probability of $2|\alpha\beta|^{2}$. Then with the help of the optical switch (OS), Bob still makes the photons in the spatial mode c$_{1}$ and b$_{1}$ pass though the $50:50$ beam splitter, here named BS$_{1}$, which can make
\begin{eqnarray}
 \hat{c}_{1}^{\dagger}|0\rangle=\frac{1}{\sqrt{2}}(\hat{d}_{1}^{\dagger}|0\rangle-\hat{d}_{2}^{\dagger}|0\rangle)\nonumber\\ \hat{b}_{1}^{\dagger}|0\rangle=\frac{1}{\sqrt{2}}(\hat{d}_{1}^{\dagger}|0\rangle+\hat{d}_{2}^{\dagger}|0\rangle).
 \end{eqnarray}

After the BS$_{1}$, the two-photon system can be described as
\begin{eqnarray}
|\psi_{1}\rangle_{a_{1}d_{1}d_{2}c_{2}}&=&\frac{\alpha\sqrt{1-t}}{\sqrt{2}}|1,1,0,0\rangle_{a_{1}d_{1}d_{2}c_{2}}+\frac{\beta\sqrt{t}}{\sqrt{2}}|0,1,0,1\rangle_{a_{1}d_{1}d_{2}c_{2}}\nonumber\\
&+&\frac{\alpha\sqrt{1-t}}{\sqrt{2}}|1,0,1,0\rangle_{a_{1}d_{1}d_{2}c_{2}}-\frac{\beta\sqrt{t}}{\sqrt{2}}|0,0,1,1\rangle_{a_{1}d_{1}d_{2}c_{2}}.
\end{eqnarray}
It can be found that if the detector D$_{1}$ fires, Eq. (\ref{after QND}) will collapse to Eq. (\ref{connection}), while if the detector D$_{2}$ fires, Eq. (\ref{after QND}) will collapse to Eq. (\ref{connection1}). There is only a phase difference between Eq. (\ref{connection}) and Eq. (\ref{connection1}), and Eq. (\ref{connection1}) can be converted to Eq. (\ref{connection}) by the phase flip operation. Then, similar to Section. II, if they can find a suitable VBS, which makes $t=\alpha^{2}$, Eq. (\ref{connection}) can evolve to Eq. (\ref{max entangle}). That is to say, Alice and Bob can distill the maximally single-photon entangled state from the arbitrary less-entangled single-photon state, with the probability $P_{1}$=2$|\alpha\beta|^{2}$.
Interestingly, it can be found when $t=\alpha^{2}$, the discarded items which make the coherent state pick up no phase shift can be rewritten as
\begin{eqnarray}
|\psi_{2}\rangle_{a_{1}b_{1}c_{1}c_{2}}&=&\alpha^{2}|1,0,0,1\rangle_{a_{1}b_{1}c_{1}c_{2}}+\beta^{2}|0,1,1,0\rangle_{a_{1}b_{1}c_{1}c_{2}},\label{discard item}
\end{eqnarray}
 with the probability of $|\alpha|^{4}+|\beta|^{4}$. Then Bob uses the OS to  make photons in the spatial modes c$_{1}$ and c$_{2}$ pass through another $50:50$ beam splitter, here named BS$_{2}$, which makes
 \begin{eqnarray}
 \hat{c}_{1}^{\dagger}|0\rangle=\frac{1}{\sqrt{2}}(\hat{e}_{1}^{\dagger}|0\rangle-\hat{e}_{2}^{\dagger}|0\rangle)\nonumber\\ \hat{c}_{2}^{\dagger}|0\rangle=\frac{1}{\sqrt{2}}(\hat{e}_{1}^{\dagger}|0\rangle+\hat{e}_{2}^{\dagger}|0\rangle).
\end{eqnarray}
After the BS$_{2}$, the discarded items can evolve to
\begin{eqnarray}
|\psi_{2}'\rangle_{a_{1}b_{1}e_{1}e_{2}}&=&\alpha^{2}|1,0,1,0\rangle_{a_{1}b_{1}e_{1}e_{2}}+\beta^{2}|0,1,1,0\rangle_{a_{1}b_{1}e_{1}e_{2}}\nonumber\\
&+&\alpha^{2}|1,0,0,1\rangle_{a_{1}b_{1}e_{1}e_{2}}-\beta^{2}|0,1,0,1\rangle_{a_{1}b_{1}e_{1}e_{2}}.
\end{eqnarray}
We can easily find if the detector E$_{1}$ fires, the Eq. (\ref{discard item}) will collapse to
\begin{eqnarray}
|\psi_{2}\rangle_{a_{1}b_{1}}&=&\alpha^{2}|1,0\rangle_{a_{1}b_{1}}+\beta^{2}|0,1\rangle_{a_{1}b_{1}},\label{new entangle}
\end{eqnarray}
while if the detector E$_{2}$ fires, the Eq. (\ref{discard item}) will collapse to
\begin{eqnarray}
|\psi_{2}'\rangle_{a_{1}b_{1}}&=&\alpha^{2}|1,0\rangle_{a_{1}b_{1}}-\beta^{2}|0,1\rangle_{a_{1}b_{1}}.\label{new entangle1}
\end{eqnarray}
 Eq. (\ref{new entangle1}) can be easily converted to Eq. (\ref{new entangle}) by the phase flip operation. Comparing with  the Eq. (\ref{singlephoton}), we can find that Eq. (\ref{new entangle}) has the similar form of Eq. (1). That is to say, Eq. (\ref{new entangle}) is a new less-entangled single-photon state and can be reconcentrated for the next round.
 In the second round of concentration, the single photon source also emits a photon and sends it to Bob. By making the photon pass through the VBS, we can create a new entangled single photon state with the form in Eq. (\ref{VBS}). Then the whole two-photon system can be written as
 \begin{eqnarray}
 &&|\phi'\rangle_{a_{1}b_{1}c_{1}c_{2}}=|\psi_{2}\rangle_{a_{1}b_{1}}\otimes|\phi\rangle_{c_{1}c_{2}}\nonumber\\
&=&\alpha^{2}\sqrt{1-t}|1,0,1,0\rangle_{a_{1}b_{1}c_{1}c_{2}}+\beta^{2}\sqrt{t}|0,1,0,1\rangle_{a_{1}b_{1}c_{1}c_{2}}\nonumber\\
&+&\alpha^{2}\sqrt{t}|1,0,0,1\rangle_{a_{1}b_{1}c_{1}c_{2}}+\beta^{2}\sqrt{1-t}|0,1,1,0\rangle_{a_{1}b_{1}c_{1}c_{2}}.\label{entangle generation2}
 \end{eqnarray}

 According to the concentration step in the first concentration round, Bob makes the photons in the spatial modes b$_{1}$ and c$_{1}$ pass through the QND, and selects the items which make the coherent state pick up $\theta$. Then the Eq. (\ref{entangle generation2}) can collapse to
 \begin{eqnarray}
 |\phi_{1}'\rangle_{a_{1}b_{1}c_{1}c_{2}}&=&\alpha^{2}\sqrt{1-t}|1,0,1,0\rangle_{a_{1}b_{1}c_{1}c_{2}}
 +\beta^{2}\sqrt{t}|0,1,0,1\rangle_{a_{1}b_{1}c_{1}c_{2}},\label{pick up}
 \end{eqnarray}
 with the probability of $P_{2}=\frac{2|\alpha\beta|^{4}}{|\alpha|^{4}+|\beta|^{4}}$, where the subscript "2" means in the second concentration round.

Then, by making the photons in the spatial modes b$_{1}$ and c$_{1}$ passing through the BS$_{1}$, Eq. (\ref{pick up}) can ultimately collapse to
\begin{eqnarray}
|\psi\rangle_{a_{1}c_{2}}=\alpha^{2}\sqrt{1-t}|1,0\rangle_{a_{1}c_{2}}+\beta^{2}\sqrt{t}|0,1\rangle_{a_{1}c_{2}}.\label{max2}
\end{eqnarray}
Here, Bob only needs to choose another suitable VBS, which makes $t_{2}=\frac{|\alpha|^{4}}{|\alpha|^{4}+|\beta|^{4}}$, where the subscript '2' means in the second concentration round. Eq. (\ref{max2}) can evolve to Eq. (\ref{max entangle}), which is the maximally single-photon entangled state. So far, we have succeeded to distill the maximally single-photon entangled state in the second concentration round, with the probability of $P_{2}=\frac{2|\alpha\beta|^{4}}{|\alpha|^{4}+|\beta|^{4}}$. Moreover, after making the photons in the modes c$_{1}$ and c$_{2}$ passing through  BS$_{2}$, the discarded items in the second concentration round can evolve to
\begin{eqnarray}
|\psi_{3}\rangle_{a_{1}b_{1}}&=&\alpha^{4}|1,0\rangle_{a_{1}b_{1}}+\beta^{4}|0,1\rangle_{a_{1}b_{1}},\label{third}
\end{eqnarray}
 which can be reconcentrated in the third round. Therefore, it can be seen that by choosing the suitable VBS with the transmission $t_{N}= \frac{|\alpha|^{2^{N}}}{|\alpha|^{2^{N}}+|\beta|^{2^{N}}}$, where the subscript "N" means the iteration number, the ECP can be used repeatedly to distill the maximally single-photon entangled  state from the less-entangled single-photon state.

\section{Discussion and summary}
 By far, we have presented two efficient ECPs for distilling the maximally entangled single-photon state from the less-entangled state.  Interestingly, in these  protocols, only one pair of nonlocal less-entangled single-photon state has to be shared by the two parties, while the auxiliary single photon is local. In Ref. \cite{shengqic}, in each concentration step, two pairs of less-entangled single-photon states have to be shared. As the nonlocal entanglement sources are expensive, our protocols are more economic.  Our first protocol is based on the linear optics, which can be realized in current
  experimental condition. In our protocols, the VBS is the key element to perform the both protocols. We need to adjust the transmission and reflection coefficients of the VBS according to the initial less-entangled state. In our second protocol, in order to repeat this ECP, we also should choose different VBS in each concentration step. The VBS is a common linear  optical element in current technology. Recently, Osorio \emph{et al.} reported their results about heralded photon amplification for quantum communication with the help of the VBS \cite{amplification}.
   They used their setup to increase the probability $\eta_{t}$ of the single photon $|1\rangle$ from a mixed state $\eta_{t}|1\rangle\langle1|+(1-\eta_{t})|0\rangle\langle0|$. In their experiment, they adjust the splitting ratio of VBS from 50:50 to 90:10 to increase
   the visibility from 46.7 $\pm$ 3.1\% to 96.3 $\pm$ 3.8\%.

In  both ECPs, the processing of the photons passing through the BS is essentially  the Hong-Ou-Mandel (HOM) interference. So the two photons should be indistinguishable in every degree of freedom. In Ref. \cite{amplification}, they have measured the  HOM interference on each BSs. Their experimental results for each BSs are 93.4 $\pm $5.9\% and 92.1 $\pm$ 5.7\%, respectively. In the first protocol, the sophisticated single photon detectors are required to  exactly  distinguish the single photon in each output modes. Current available detectors are InGaAs/InP avalanche photodiodes. Fortunately, available single photon detector with 10\% efficiency, 3 KHz noise and 20 $\mu$s dead time in free running are suitable for experiment.
In fact, the setup of our first ECP is quite analogy
with the theoretical setup in Ref.\cite{amplification}. In their setup, they choose the VBS with $t>\frac{1}{2}$  to realize
the photon amplification while we choose $t=\alpha^{2}$ to complete the concentration. Therefore,  their
experimental setup can also be used to perform our first ECP by choosing different VBSs.

In the second ECP, the most important element is the cross-Kerr nonlinearity. In the practical applications, the cross-Kerr nonlinearity has been regarded as a controversial topic for a long time \cite{Banacloche,Shapiro1,Shapiro2}, for during the homodyne detection process, decoherence is inevitable, which may lead the qubit states degrade to the mixed states \cite{decoherent1,decoherent2}. Meanwhile,  the natural cross-Kerr nonlinearity is extremely weak so that it is difficult to determine the phase shift due to the impossible discrimination of two overlapping coherent states in homodyne detection \cite{purification2}. Fortunately, according to Ref. \cite{decoherent1}, we can make the decoherence extremely weak, simply by an arbitrary strong coherent state associated with a displacement D($-\alpha$) performed on the coherent state. Moreover, several theoretical works have proved that with the help of
weak measurement, it is possible for the phase shift to reach an observable value \cite{lin1,lin2,weak_meaurement,oe}.

Finally, it is interesting to calculate the total success probability of the second ECP. According to the concentration steps in Sec.III, we can obtain the success probability in each concentration round, which can be written as
\begin{eqnarray}
P_{1}&=&2|\alpha\beta|^{2},\nonumber\\
P_{2}&=&\frac{2|\alpha\beta|^{4}}{|\alpha|^{4}+|\beta|^{4}},\nonumber\\
P_{3}&=&\frac{2|\alpha\beta|^{8}}{(|\alpha|^{4}+|\beta|^{4})(|\alpha|^{8}+|\beta|^{8})},\nonumber\\
P_{4}&=&\frac{2|\alpha\beta|^{16}}{(|\alpha|^{4}+|\beta|^{4})(|\alpha|^{8}+|\beta|^{8})(|\alpha|^{16}+|\beta|^{16})},\nonumber\\
&\cdots\cdots&\nonumber\\
P_{N}&=&\frac{2|\alpha\beta|^{2^{N}}}{(|\alpha|^{4}+|\beta|^{4})(|\alpha|^{8}+|\beta|^{8})\cdots(|\alpha|^{2^{N}}+|\beta|^{2^{N}})^{2}}.\label{probability}
\end{eqnarray}

\begin{figure}[!h]
\begin{center}
\includegraphics[width=8cm,angle=0]{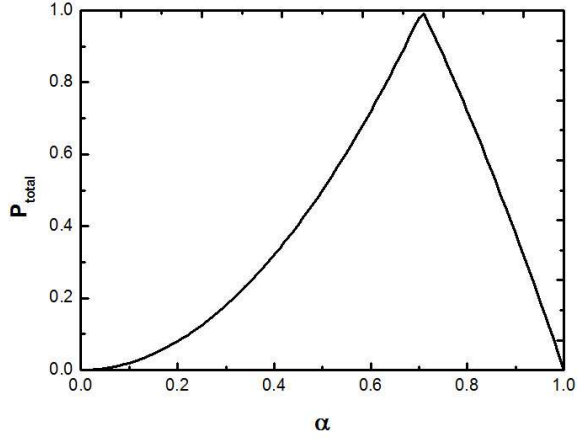}
\caption{The success probability ($P_{total}$) for obtaining a maximally single-photon entangled state after the concentration protocol being operated for N times. For numerical simulation, we choose $N=10$. It can be seen that the value of $P_{total}$ largely depends on the initial coefficient $\alpha$. When $\alpha=\frac{1}{\sqrt{2}}$, $P_{total}$ reaches the maximum as 1.}
\end{center}
\end{figure}
In theory, the second ECP can be reused indefinitely, the total success probability for distilling the maximally single-photon entangled state from the less-entangled single-photon state equals the sum of the success probability in each concentration round.
\begin{eqnarray}
P_{total}=P_{1}+P_{2}+\cdots P_{N}=\sum\limits_{N=1}^{\infty} P_{N}.
\end{eqnarray}
According to Eq. ({\ref{probability}}), we can find that if the original state is the maximally single-photon entangled state, where $\alpha=\beta=\frac{1}{\sqrt{2}}$, the probability $P_{total}=\frac{1}{2}+\frac{1}{4}+\frac{1}{8}+\cdots\frac{1}{2^{N}}+\cdots=1$, while if the original state is the less-entangled single-photon state, where $\alpha\neq\beta$, the $P_{total}<1$. Here, we choose $N=10$ as a proper approximation and calculate the value of $P_{total}$ in different original entanglement coefficient. Fig. 3 shows the value of $P_{total}$ as a function of the entanglement coefficient $\alpha$. It can be found that $P_{total}$ largely depends on the original entanglement state. The main reason is that this  ECP is based on the LOCC. It is well known that the LOCC cannot increase the entanglement. Therefore, the essence of the entanglement concentration is the entanglement transformation. The entanglement of the concentrated state comes from the initial less-entangled state. If  the entanglement of the  initial less-entangled state is low, the total success probability for obtaining the maximally entangled state is also small.

In conclusion, we have put forward two efficient ECPs for distilling the maximally single-photon entangled state with an auxiliary single photon. In these two ECPs, the auxiliary photon is only possessed by Bob. Bob can operate all the concentration steps alone and tell Alice the measurement results, which can simplify the experimental operation largely. The first ECP is with the linear optics and can be realized in current experimental conditions. In this ECP, with the help of Bell measurement, we can obtain the maximally single-photon entangled state with the success probability of $P=2|\alpha\beta|^{2}$. In the second ECP, the weak cross-Kerr nonlinearity is used to complete the nondestructive photon number detection. Comparing with other ECPs, the second ECP has several advantages: first, it does not require the sophisticated single photon detectors, ordinary photon detectors can also achieve this protocol; second, after the photon number detection, the photons can be remained for the next step; third, by choosing suitable VBS, the second ECP can be used repeatedly to get a higher success probability; forth, only Bob needs to operate the whole step. Therefore, our ECPs, especially the second ECP may be useful and convenient in the current quantum information processing.

\section*{ACKNOWLEDGEMENTS}
 This work is supported by the National Natural Science Foundation of
China under Grant No. 11104159, Open Research
Fund Program of the State Key Laboratory of
Low-Dimensional Quantum Physics Scientific, Tsinghua University,
Open Research Fund Program of National Laboratory of Solid State Microstructures under Grant No. M25020 and M25022, Nanjing University,  and the Project Funded by the Priority Academic Program Development of Jiangsu
Higher Education Institutions.


\begin{thebibliography}{10}

\bibitem{book} Nielsen M A and Chuang I L 2000 {\it Quantum Computation and
Quantum Information} (Cambridge University Press, Cambridge,
England)

\bibitem{rmp} Gisin N, Ribordy G, Tittel W and Zbinden H 2002
{\it Rev. Mod. Phys.} {\bf 74} 145

\bibitem{teleportation1} Bennett C H, Brassard G, Crepeau C, Jozsa R, Peres A and Wootters W K 1993 {\it Phys. Rev. Lett.} {\bf 70} 1895

\bibitem{densecoding1} Bennett C H  and  Wiesner S J 1992 {\it  Phys. Rev. Lett.} {\bf 69} 2881




\bibitem{QSS1} Hillery M, Bu\v{z}ek V and Berthiaume A  1999 {\it Phys. Rev. A} {\bf 59} 1829

\bibitem{QSS2} Karlsson A, Koashi M  and Imoto N 1999 {\it Phys. Rev. A} {\bf 59}
162

\bibitem{QSS3} Xiao L, Long G L, Deng F G and Pan J W  2004 {\it Phys. Rev. A} {\bf 69}  052307



\bibitem{Ekert91}  Ekert A K  {\it Phys. Rev. Lett.} 1991  {\bf 67} 661

\bibitem{BBM92} Bennett C H,  Brassard G  and  Mermin N D  1992 {\it Phys. Rev. Lett.}
{\bf 68} 557


\bibitem{long}  Long G L and  Liu X  S 2002  {\it Phys. Rev. A} {\bf 65} 032302


\bibitem{two-step}Deng F G,  Long G L and Liu X S 2003  {\it Phys. Rev.
A} {\bf 68} 042317 .

\bibitem{lixhpra}  Li X H,  Deng F G and  Zhou H Y 2006 {\it Phys. Rev. A}
{\bf 74} 054302

\bibitem{QKDdeng1} Deng F G  and  Long G L {\it Phys. Rev. A} 2003 {\bf 68} 042315

\bibitem{lombardi} Lombardi E, Sciarrino F, Popescu S and Martini F De 2002
{\it Phys. Rev. Lett.} {\bf 88} 070402

\bibitem{chou} Chou C W , Laurat J, Deng H, Choi K S, Riedmatten H, Felinto D, Kimble H J2007 {\it Science} {\bf 316} 1316

\bibitem{memory}Duan L M, Lukin M D, Cirac J I and Zoller P 2001 {\it Nature} {\bf 414}, 413


\bibitem{singlephotonrepeater3} Simon C, Riedmatten H De, Afzelius M,
Sangouard N, Zbinden H and Gisin N 2007 {\it Phys. Rev. Lett.} {\bf 98}
0190503

\bibitem{telescope} Gottesman D, Jennewein T,and Croke S, 2012 {\it Phys. Rev. Lett.} {\bf 109}, 070503
\bibitem{repeater1} Min$\acute{a}$$\breve{r}$ J, Riedmatten H De, Simon C, Zbinden H and Gisin N 2008 {\it Phys. Rev. A} {\bf 77} 052305

 \bibitem{sangouard} Sangouard N, Simon C, Coudreau T and Gisin N 2008 {\it Phys. Rev. A} {\bf 78} 050301(R)

\bibitem{sangouard2}Salart D, Landry O,Sangouard N,  Gisin N, Herrmann  H, Sanguinetti B,
Simon C, Sohler W, Thew R T, Thomas A, and Zbinden H  2012 {\it Phys. Rev. Lett.} {\bf 104}180504


\bibitem{shengqic}  Sheng Y B, Deng F G and Zhou H Y 2010 {\it Quant.
Inf. \& Comput.} {\bf 10} 272



\bibitem{C.H.Bennett2} Bennett C H, Bernstein H J, Popescu S and
Schumacher B 1996 {\it Phys. Rev. A} {\bf 53} 2046

\bibitem{swapping1} Bose S, Vedral V  and Knight P L 1999 {\it Phys. Rev. A}
{\bf 60} 194
\bibitem{swapping2} Shi B S, Jiang Y K and Guo G C 2000 {\it Phys.
Rev. A} {\bf 62} 054301



\bibitem{zhao1} Zhao Z, Pan J W and Zhan M S 2001 {\it Phys. Rev. A} {\bf 64} 014301

\bibitem{Yamamoto1} Yamamoto T, Koashi M and Imoto N 2001 {\it Phys. Rev.
A} {\bf 64} 012304



\bibitem{wangxb} Wang X B and Fan H 2003 {\it Phys. Rev. A} {\bf 68}
060302

\bibitem{shengpra2} Sheng Y B, Deng F G and Zhou H Y 2008 {\it Phys.
Rev. A} {\bf 77} 062325

\bibitem{shengpra3} Sheng Y B, Zhou L, Zhao S M and
Zheng B Y 2012 {\it Phys. Rev. A} {\bf 85} 012307

\bibitem{shengpra4}Sheng Y B, Zhou L and Zhao S M 2012  {\it Phys. Rev. A} {\bf 85} 044305

\bibitem{dengpra} Deng F G 2012 {\it Phys. Rev. A} {\bf 85} 022311

\bibitem{wangc} Wang C 2012  {\it Phys. Rev. A} {\bf 86}, 012323

\bibitem{kok1} Kok P, Munro W J, Nemoto K, Ralph T C, Dowing J P and Milburn G J 2007 {\it Rev. Mod. Phys.} {\bf 79} 135

\bibitem{kok2} Kok P, Lee H and Dowling J P 2002 {\it Phys. Rev. A} {\bf 66} 063814


\bibitem{detection1} Gao M, Hu W H and Li C Z 2007 {\it J. Phys. B: At. Mol. Opt. Phys.} {\bf 40} 3525

\bibitem{gate} Nemoto K and Munro W J 2004 {\it Phys. Rev. Lett.} {\bf 93} 250502
\bibitem{gate1} Munro W J, Nemoto K and Spiller T P 2005 {\it New J. Phys.} {\bf 7} 137


\bibitem{QND2} Barrett S D, Kok P, Nemoto K, Beausoleil R G, Munro W J
 and Spiller T P 2005 {\it Phys. Rev. A} {\bf 71} 060302
\bibitem{shengbellstateanalysis} Sheng Y B, Deng F G and
Long G L 2010 {\it Phys. Rev. A} {\bf 82} 032318


\bibitem{qi} Guo Q, Bai J, Cheng L Y, Shao X Q, Wang H F and
Zhang S 2011 {\it Phys. Rev. A} {\bf 83} 054303

\bibitem{he1} He B, Bergou J A and Ren Y H 2007 {\it Phys. Rev. A} {\bf 76}, 032301


\bibitem{he2} He B, Lin Q and Simon C 2011 {\it Phys. Rev. A} {\bf 83} 053826

\bibitem{he3}He B, Ren Y H and Bergou J A 2009 {\it J. Phys. B} {\bf 43}
025502

\bibitem{lin1} Lin Q and Li J 2009 {\it Phys. Rev. A} {\bf 79} 022301

\bibitem{lin2} Lin Q and He B 2009 {\it Phys. Rev. A} {\bf 80} 042310

\bibitem{lin3} Lin Q, He B,
Bergou J A and Ren Y H 2009  {\it Phys. Rev. A} {\bf 80} 042311


\bibitem{xiayan1}Xia Y, Song J, Lu P M and Song H S 2011  {\it J. Phys. B: At. Mol. Opt. Phys.} {\bf 44} 079804

\bibitem{xiayan2}Xia Y, Song J, Lu P M and Song H S 2011  {\it J. Phys. B: At. Mol. Opt. Phys.} {\bf 44} 025503

\bibitem{zhangzhiming}Zhang Z M, Yang J and Yu Y F 2008 {\it J. Phys. B: At. Mol. Opt. Phys.} {\bf 41} 025502

\bibitem{guoyu}Guo Y and Kuang L M 2007 {\it J. Phys. B: At. Mol. Opt. Phys.} {\bf 40} 3309

\bibitem{amplification} Osorio C I, Bruno N, Sangouard N, Zbinden H, Gisin N and Thew R T 2012 {\it Phys. Rev. A}  {\bf 86}  023815


\bibitem{Banacloche} Nielsen A E B, Muschik C A, Giedke G and Vollbrecht K G H 2010 {\it Phys. Rev. A} {\bf 81} 043823

\bibitem{Shapiro1} Shapiro J H 2006 {\it Phys. Rev. A} {\bf 73} 062305

\bibitem{Shapiro2} Shapiro J H and Razavi M 2007 {\it New J. Phys.} {\bf 9} 16

\bibitem{decoherent1} Jeong H 2006 {\it Phys. Rev. A} {\bf 73} 052320
\bibitem{decoherent2} Barrett S D and Milburn G J 2006 {\it Phys. Rev. A} {\bf 74} 060302(R)

\bibitem{purification2} Wang C, Zhang Y and Jin G S 2011 {\it J. Mod. Optics} {\bf 58} 21

\bibitem{weak_meaurement} Feizpour A, Xing X and Steinberg A M {\it Phys. Rev. Lett.} {\bf 107}
    133603

\bibitem{oe} Zhu C and Huang G 2011 {\it Optics Express} {\bf 19} 23364

\end{thebibliography}
\end{document}